\documentclass[tightenlines,showpacs,prx,floatfix,preprintnumbers,amsmath,superscriptaddress,
amssymb,amsbsy,nofootinbib,nobibnotes,twocolumn,longbibliography,prx]{revtex4-1}
\usepackage{color}
\usepackage{graphicx}
\usepackage{natbib}
\usepackage{verbatim}
\usepackage{soul}

\newcommand{\blue}[1]{#1}
\newcommand{\yr} {
  \,\mathrm{yr}
  }
\newcommand{\Msun}{\,\mathrm{M}_\odot}
\setstcolor{blue}

\begin{document}

\title{Observing Gravitational Waves with a Single Detector}
\author{T. A. Callister}
\email{tcallist@caltech.edu}
\author{J. B. Kanner}
\author{T. J. Massinger}
\affiliation{LIGO Laboratory, California Institute of Technology, Pasadena, CA 91125, USA}
\author{S. Dhurandhar}
\affiliation{Inter-University Centre for Astronomy and Astrophysics, Pune 411007, India}
\author{A. J. Weinstein}
\affiliation{LIGO Laboratory, California Institute of Technology, Pasadena, CA 91125, USA}

\begin{abstract}
A major challenge of any search for gravitational waves is to distinguish true astrophysical signals from those of terrestrial origin.
Gravitational-wave experiments therefore make use of multiple detectors, considering only those signals which appear in coincidence in two or more instruments.
It is unclear, however, how to interpret loud gravitational-wave candidates observed when only one detector is operational.
In this paper, we demonstrate that the observed rate of binary black hole mergers can be leveraged in order to make confident detections of gravitational-wave signals with one detector alone.
We quantify detection confidences in terms of the probability $P(S)$ that a signal candidate is of astrophysical origin.
We find that, at current levels of instrumental sensitivity, loud \blue{binary black hole} candidates observed with a single Advanced LIGO detector can be assigned $P(S)\gtrsim0.4$.
In the future, Advanced LIGO may be able to observe \blue{binary black hole mergers with single-detector confidences} exceeding $P(S)\sim90\%$.
\end{abstract}

\preprint{Published As: Class. Quantum Grav. \textbf{34}, 155007 (2017)
\hspace{2.9cm}
DOI: https://doi.org/10.1088/1361-6382/aa7a76}

\maketitle

\section{Introduction}

Distinguishing true gravitational wave signals from local disturbances is a basic challenge for all terrestrial gravitational wave observatories.
In the case of interferometers like Advanced LIGO \cite{Aasi2015,Abbott2016d}, Advanced Virgo \cite{Acernese2015}, and KAGRA \cite{Aso2013}, noise transients may arise from a long list of sources, including ground motion, power line fluctuations, magnetic fields, acoustic couplings, and non-linear mechanical motion of instrument components \cite{Abbott2016h,Vajente2016}.
To reduce the number of false-positives due to such transients, existing searches for gravitational waves rely on the simultaneous operation of multiple gravitational-wave detectors.
By requiring candidate signals to appear in coincidence in two or more detectors and through the use of the time-slide method \cite{Abbott2016i,Usman2016}, present-day searches can detect gravitational waves with extraordinary statistical confidence.
For instance, the false alarm rate (FAR) of events as significant as the binary black hole merger GW150914 is $<6.0\times10^{-7}\yr^{-1}$ in current search pipelines \cite{LIGO2016a,Abbott2016}.

Although the requirement that a signal be observed in coincidence enables high-confidence detections, this requirement is in tension with the reality of operating the current generation of detectors.
The Advanced LIGO detectors, for example, rarely achieve operational duty cycles higher than 70\%; the duty cycle is often much lower when attending to routine instrumental maintenance and repair \cite{Abbott2016d}.
Thus, for a significant fraction of the time, the gravitational-wave sky is observed with only a single detector.
Under current analysis schemes, this single-detector time cannot be meaningfully searched for astrophysical signals without independent multi-messenger confirmation from electromagnetic or particle observatories \cite{LIGOScientificCollaboration2016,Adrian-Martinez2016}.

As long as this single-detector time is discarded, the full scientific potential of existing gravitational-wave experiments will be unrealized.
A network of two detectors, each operating with uncorrelated $70\%$ duty cycles, will accumulate nearly as much single-detector observation time as coincident time (note that detector duty cycles \textit{are} in reality somewhat correlated due to seasonal weather, day/night cycles, and maintenance schedules).
Assuming that the sensitive range of the two-detector network is $\sim\sqrt{2}$ farther than the range of a single instrument, the net \textit{time-volume} observed in single-detector time is related to that probed in coincidence by
	\begin{equation}
	\frac{\langle VT\rangle_\mathrm{single}}{\langle VT\rangle_\mathrm{coinc}} \sim 0.30
	\end{equation}
Thus, for every three \blue{binary black hole mergers} observed by Advanced LIGO in coincidence, we might expect approximately one additional event in single-detector time.
Moreover, there is a \blue{non-negligible chance} that the next gravitational-wave signal of profound importance (whether a signal with high signal-to-noise ratio, unusual masses, or even the first observed binary neutron star merger) will arrive when only one detector is operational.
\blue{The gravitational-wave community will need a means for quantifying the significance of such events.}

In this paper, we present a framework for assigning significance to gravitational-wave events observed with only one detector.
By leveraging the measured rate of binary black hole mergers \cite{Abbott2016f,Abbott2016g,Abbott2016}, we demonstrate that loud gravitational-wave candidates in single-detector time can be assigned high probabilities of astrophysical origin.
We note that ours is not the first framework to accommodate the ranking of single-detector triggers \cite{Cannon2015,Messick2017}.
Earlier efforts estimate the false alarm rate (FAR) of signal candidates in single-detector time but have stopped short of considering probabilities of astrophysical origin.

\section{Calculating Astrophysical Probabilities}
\label{calc_confidence}

Gravitational-wave candidates (or ``triggers'') arise from two distinct populations:
the population of true astrophysical gravitational-wave signals, and the population of terrestrial artifacts due to detector noise or environmental effects.
We will assume that the signal and noise populations each obey Poisson statistics.
Given a gravitational-wave candidate with detection statistic $\rho$, the probability that the candidate is a true astrophysical signal is \cite{Farr2015,Abbott2016f,Abbott2016}
	\begin{equation}
	P(S|\Lambda_s,\Lambda_n) = \frac{\Lambda_s p_s(\rho)}{\Lambda_s p_s(\rho) + \Lambda_n p_n(\rho)},
	\label{pAstroIntegrand}
	\end{equation}
where $p_s(\rho)$ and $p_n(\rho)$ are the probability densities describing the distribution of detection statistics $\rho$ under each population.
The densities $p_s(\rho)$ and $p_n(\rho)$ are normalized on the interval $\rho\in(\rho_\text{min},\rho_\text{max})$, where $\rho_\text{min}$ and $\rho_\text{max}$ are the minimum and maximum detection statistics considered in the search.
In our analysis below, we will take $\rho_\mathrm{min}=7.1$ and $\rho_\mathrm{max}=\infty$, unless otherwise noted.
Meanwhile, $\Lambda_s$ and $\Lambda_n$ are the mean Poisson rates of signal and noise triggers with $\rho>\rho_\mathrm{min}$.
The rates $\Lambda_s$ and $\Lambda_n$ are not precisely known, of course, and in practice we will marginalize over these quantities, giving
	\begin{equation}
	P(S) = \int  \frac{\Lambda_s p_s(\rho)}{\Lambda_s p_s(\rho) + \Lambda_n p_n(\rho) }
		p(\Lambda_s) p(\Lambda_n) d\Lambda_s d\Lambda_n,
	\label{pAstro}
	\end{equation}
where $p(\Lambda_s)$ and $p(\Lambda_n)$ are the priors placed on each rate.

Eq. \eqref{pAstro}, as well as our methodology discussed below, may be generalized to any detection statistic $\rho$.
For concreteness, though, in this paper we will specialize to the re-weighted signal-to-noise ratio (SNR) statistic adopted in the \texttt{PyCBC} analysis pipeline \cite{Canton:2014ena,Usman2016,pycbc}.
\blue{
The re-weighted SNR combines the standard matched-filter SNR with a chi-squared statistic quantifying the match between an observed candidate and its best-fit template waveform.
For a true astrophysical signal, the re-weighted SNR approximately equals the signal's standard matched-filter SNR.
For instrumental glitches and terrestrial artifacts with large chi-squared values, the re-weighted SNR statistic is downgraded appropriately.
Hereafter, we will use the variable $\rho$ to refer to a re-weighted SNR.
}

The signal-to-noise ratio with which a gravitational-wave event is detected scales as $\mathrm{SNR}\propto D^{-1}$, where $D$ is the distance to the source.
Assuming that gravitational-wave events in the local Universe are distributed uniformly in volume, the probability density of astrophysical triggers follows $p_s(\rho)\propto \rho^{-4}$.
\blue{Note that $p_s(\rho)$ will receive corrections due to signals at non-negligible redshifts; for simplicity we will ignore these corrections here.
}
Meanwhile, the ``background'' distribution $p_n(\rho)$ of noise triggers is determined experimentally by measuring the frequency of noise events in the gravitational-wave detectors.
This is conventionally done via the time-slide method (or other related schemes), in which a null stream is constructed by temporally shifting data from two detectors with respect to one another \cite{Abbott2016i,Usman2016,Cannon2015,Messick2017}.

The time-slide method, however, cannot be used to form a background for single-detector events.
Instead, we will construct a single-detector background using triggers that fall in \textit{coincident} time: times in which two or more gravitational-wave detectors are operational.
Specifically, we will construct $p_n(\rho)$ for a given detector by selecting all triggers (a) observed by the detector during coincident time but which (b) did not occur in coincidence with a trigger at another site.
These triggers likely do not represent true gravitational-wave signals as they were not observed in multiple detectors, and so they must instead arise from the population of noise events.

\begin{figure}
\centering
\includegraphics[width=0.48\textwidth]{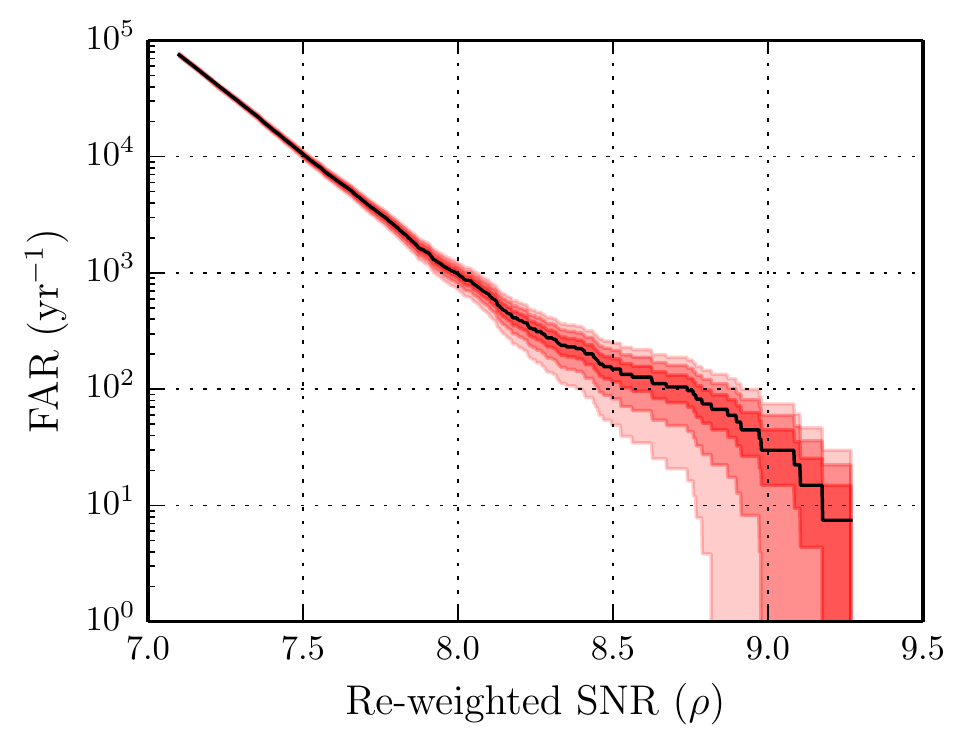} 
\caption{
The \blue{false alarm rate} of H1 single-detector triggers \blue{above a given $\rho$} during Advanced LIGO's O1 observing run.
The background shown corresponds to the \texttt{PyCBC} search region (ii) \cite{Abbott2016,Canton:2014ena,Usman2016,pycbc}.
All triggers occurred in times during which L1 was operating, but were not observed in coincidence with a trigger in L1.
The set of triggers shown is therefore a measure of noise events specific to H1.
The shaded regions show one-, two-, and three-sigma Poisson uncertainties on the measured rate of noise events.
In this example, the loudest background event has detection statistic $\rho_\mathrm{thresh} = 9.3$ with a false alarm rate $\mathrm{FAR}=7.4\yr^{-1}$.
}
\label{H1Background}
\end{figure}

As an example of this procedure, we show in Fig. \ref{H1Background} the single-detector background constructed for Advanced LIGO's Hanford detector (H1) during the O1 observing run.
\blue{Specifically, we show the false alarm rate of single-detector candidates above a given $\rho$, obtained by the integral}
	\begin{equation}
	\blue{\mathrm{FAR}(\rho) = \int_\rho^{\rho_\mathrm{\max}}  \Lambda_n p_n(\rho') d\rho'.}
	\end{equation}
The data shown was collected over $T=0.13$ years of coincident observing time between the LIGO Hanford and Livingston (L1) detectors, and represents all H1 triggers from the \texttt{PyCBC} search region (ii) (see Appendix \ref{bulkAppendix}) \cite{Abbott2016,Abbott2016h,Canton:2014ena,Usman2016,pycbc} that were detected when L1 was operational, but which did not have a coincident L1 trigger.
This background may now be used to evaluate the significance of gravitational-wave candidates occurring when only H1 (but not L1) is operational.

From Fig. \ref{H1Background}, we see that the H1 single-detector background falls exponentially between $7.1<\rho\lesssim8.0$, consistent with Gaussian noise.
Above $\rho\approx8.0$, however, we encounter an elevated ``tail'' of non-Gaussian, high-SNR noise events.
The highest measured background event occurs at $\rho_\mathrm{thresh}=9.3$ with a corresponding false alarm rate of $7.4\yr^{-1}$.

By excluding from the background all H1 triggers observed in coincidence with a trigger in L1, we are likely underestimating the true background.
It is probable that some excluded events are, in fact, noise triggers that are in accidental coincidence with a noise event in L1.
However, we will see below that single-detector searches are sensitive to only the loudest events with $\rho\gtrsim9$, where the probability of accidental coincidence is negligible.
We therefore expect the H1 background in Fig. \ref{H1Background} to be accurate in the region of interest.

To compute the probability $P(S)$ that a single-detector event is astrophysical, we will also need prior distributions $p(\Lambda_s)$ and $p(\Lambda_n)$ on the rates of signal and noise events in the instrument.
The prior $p(\Lambda_n)$ is obtained by assuming standard Poisson uncertainty on the total rate of measured background events.
\blue{The astrophysical rate prior $p(\Lambda_s)$, meanwhile, is available for binary black hole mergers through direct Advanced LIGO observations.
There are large uncertainties, however, on the merger rate of other objects like binary neutron stars.
In this paper, we will therefore focus on the case of binary black hole signal candidates, constructing $p(\Lambda_s)$ using the measured rate of binary black hole mergers from Advanced LIGO's O1 observing run.}
Specifically, the O1 analysis yields a posterior on the rate density $R$ (rate per unit volume) of such mergers \cite{Abbott2016}.
We convert this to a posterior on the rate of measurable single-detector events using $\Lambda_s = R\langle V\rangle_\mathrm{single}$, where $\langle V\rangle_\mathrm{single}$ is the population-averaged volume inside of which binary black holes are observed with $\rho>\rho_\mathrm{min}$ in a single detector \cite{Abbott2016f,Abbott2016g}; see Appendix \ref{sec:volume} for details.

The effective live-time of our single-detector background measurement is necessarily comparable to the real amount of coincident observation time (approximately several months).
Hence the most stringent false alarm rate we can assign to a loud single-detector gravitational-wave candidate in this example is $\mathrm{FAR}\lesssim 1\,\mathrm{month}^{-1}$.
The time-slide method, in contrast, can effectively construct millions of coincident background realizations and is capable of assigning false alarm rates as low as $\mathrm{FAR}\lesssim10^{-7}\yr^{-1}$, as in the case of GW150914.
We will see, though, that a single-detector event with a marginal false alarm rate can nonetheless be assigned a strong probability $P(S)$ of astrophysical origin.

\section{Astrophysical Probabilities at High SNR}
\label{sec:beyond}

\begin{figure*}
\centering
\includegraphics[width=0.48\textwidth]{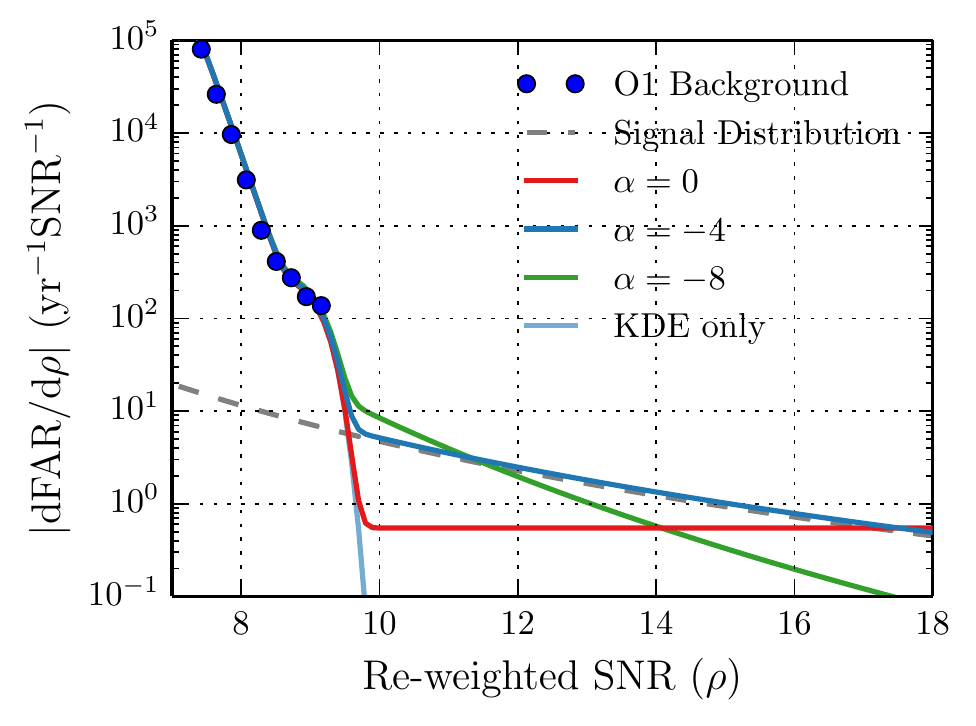}\hspace{0.5cm}
\includegraphics[width=0.48\textwidth]{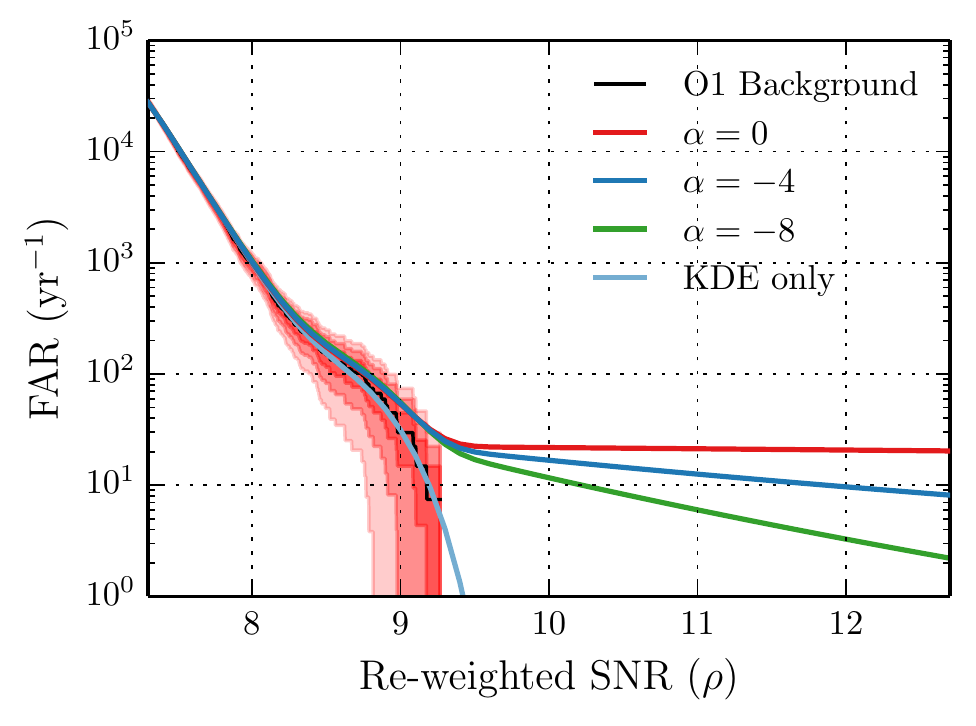}
\caption{
\textit{Left}:
The measured \blue{differential rate} $\Lambda_n p_n(\rho)$ of H1 single-detector triggers during Advanced LIGO's O1 observing run (blue points), extended with several possible models for the probability density $p_n(\rho)$.
Specifically, we consider a flat background model (red), as well as power-law models $p_n(\rho)\propto\rho^{-4}$ (blue) and $p_n(\rho)\propto\rho^{-8}$ (green).
Each of these background models is independently normalized following Eq. \eqref{lambda_n} and then summed with a \blue{Gaussian} kernel density estimation of the measured H1 background.
Also shown is the \blue{Gaussian} KDE result itself (light blue), which falls exponentially with $\rho$.
For reference, we include the expected distribution $\Lambda_s p_s(\rho)$ of astrophysical signals (dashed grey), marginalized over the measured rate $\Lambda_s$ of binary black hole mergers.
\textit{Right}:
The \blue{false alarm rate of single-detector events above a given $\rho$}, for each of the background models considered.
As in Fig. \ref{H1Background}, the black curve marks the measured H1 background during Advanced LIGO's O1 observing run, while the shaded red regions show one, two, and three-sigma Poisson uncertainties on the background rate.
The sharp drops in the left-hand plot between the measured background distribution and each of the power-law models correspond to the appearance of high-SNR tails in the cumulative rate of noise events.
}
\label{many_models}
\end{figure*}

\begin{figure}
\centering
\includegraphics[width=0.48\textwidth]{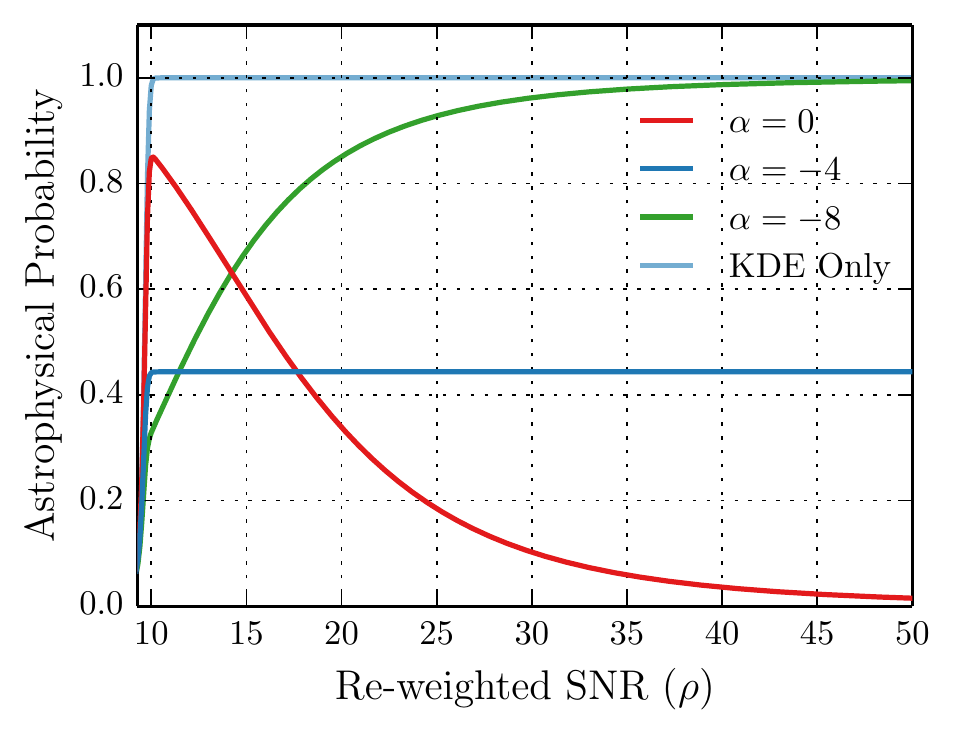}
\caption{
The astrophysical probability $P(S)$ of a gravitational-wave candidate as a function of its re-weighted SNR $\rho$, for each background model presented in the left subplot.
The KDE-only and $\rho^{-8}$ background models each fall more rapidly than the signal distribution, and hence yield $P(S)\to1$ as $\rho\to\infty$.
The flat background, meanwhile, paradoxically yields \textit{decreasing} $P(S)$ with increasing $\rho$.
Finally, the $\rho^{-4}$ model exactly parallels the expected signal distribution and therefore predicts constant $P(S)$ at large $\rho$.
}
\label{pAstro_figure}
\end{figure}

Using the measured background shown in Fig. \ref{H1Background} together with Eq. \eqref{pAstro}, we can calculate the probability $P(S)$ that a trigger falling in H1 single-detector time is astrophysical, provided $\rho<9.3$.
If instead a loud trigger is observed with $\rho>9.3$, we would still like to apply \eqref{pAstro} to determine its probability of astrophysical origin.
However, because such a trigger falls beyond our measured background, it is not obvious how to do this.
We might hope, though, to place a sensible \textit{lower limit} on $P(S)$ for such an event.

\subsection{The Naive Estimate}

As a first attempt, we outline a back-of-the-envelope prescription by which to assign a limiting $P(S)$ to a loud candidate lying above the background.
To exactly calculate a trigger's probability of astrophysical origin would require a model for the probability density $p_n(\rho)$ of noise events above $\rho_\mathrm{thresh}$.
The only experimental constraint on $p_n(\rho)$ in this region is the fact that \textit{no} noise triggers were observed above $\rho_\mathrm{thresh}$.
This fact allows us to limit the total rate $\lambda_n$ of noise triggers above $\rho_\mathrm{thresh}$, where
	\begin{equation}
	\lambda_n = \Lambda_n \int_{\rho_\mathrm{thresh}}^{\rho_\mathrm{max}} p_n(\rho) d\rho.
	\label{lambda_n}
	\end{equation}
In particular, if the noise population obeys Poisson statistics, then $\lambda_n < 3.0/T$ at 95\% credibility.
Meanwhile, the expected rate of astrophysical signals above $\rho_\mathrm{thresh}$ is
	\begin{equation}
	\lambda_s = \Lambda_s \int_{\rho_\mathrm{thresh}}^{\rho_\mathrm{max}} p_s(\rho) d\rho.
	\end{equation}
Then a simple estimate of the probability that a trigger falling above $\rho_\mathrm{thresh}$ is astrophysical is
	\begin{equation}
	P_\mathrm{Naive}(S|\rho>\rho_\mathrm{thresh}) \gtrsim \frac{\lambda_s}{\lambda_s + \lambda_n}.
	\label{naive}
	\end{equation}
	
This lower limit is sensible in the absence of any additional knowledge about the candidate signal in question (e.g. its specific detection statistic $\rho$).
Nonetheless, because we do not actually know the density $p_n(\rho)$ of noise events beyond our measured background, we might suspect that the specific measured value of $\rho$ does not provide much additional information.
Eq. \eqref{naive} is therefore likely to be a reasonable estimate of our detection confidence.
If this is indeed the case, we should expect our more careful calculation below to yield results similar to our naive expectation here.

\subsection{The Complete Calculation}

To more carefully compute $P(S)$ for triggers lying beyond the background, we must adopt a specific model for the background density $p_n(\rho)$ above $\rho_\mathrm{thresh}$.
We have a great deal of freedom in this choice.
While our model must reproduce the normalization constraint on $\lambda_n$ [see Eq. \eqref{lambda_n}], there are no other \textit{a priori} restrictions on the \textit{shape} of the model.

\blue{In the left side of Fig. \ref{many_models}, the blue points show the differential rate $|d\mathrm{FAR}/d\rho| = \Lambda_n p_n(\rho)$ of H1 single-detector background triggers in O1.
This measured background is joined with several possible models for the background background at high SNR}:
a flat model with $p_n(\rho)=\mathrm{constant}$ (red), and power law models $p_n(\rho) \propto \rho^\alpha$ with $\alpha = -4$ (blue) and $\alpha= -8$ (green).
The amplitude of each model is fixed by the normalization condition $\lambda_n = 3/T$.
While the power-law models are normalized using $\rho_\mathrm{max} = \infty$, for the flat background model we arbitrarily set $\rho_\mathrm{max}=50$, as the flat model is otherwise unnormalizable.
After normalization, each model is added to a \blue{Gaussian} kernel density estimation (KDE) of the measured background to obtain a smooth distribution between $\rho_\mathrm{min}$ and $\rho_\mathrm{max}$.
Also included in Fig. \ref{many_models} is a Gaussian background model (light blue) obtained through kernel density estimation of the measured noise triggers alone; this model \textit{does not} obey the normalization condition above.
The dashed grey curve shows the inferred distribution of astrophysical signals, obtained after marginalization over $\Lambda_s$.

Note that there is some ambiguity in the choice of KDE bandwidth used in Fig. \ref{many_models}; different reasonable choices can lead to factor of $\sim2$ differences in the background height between $\rho\approx9$ and $10$.
This uncertainty, however, is much smaller than the uncertainty in the binary black hole merger rate $\Lambda_s$.

Each model in the left-hand side of Fig. \ref{many_models} exhibits a sharp drop at $\rho\approx9.5$.
While initially unsettling, such drops in fact correspond to the appearance of a high-SNR tail in the cumulative background rate.
To illustrate this, the right-hand side of Fig. \ref{many_models} shows the cumulative \blue{false alarm rate of single-detector triggers under each model}, together with the measured H1 background from Fig. \ref{H1Background}.
With the exception of the ``KDE only" model, each background model yields a kink in the cumulative rate at $\rho\approx9.5$, transitioning into a tail at high-SNR.
\blue{To understand this behavior, note that the differential rate $\Lambda_n p_n(\rho)$ is the (negative) derivative of the total false alarm rate.
A sharp drop in the differential rate therefore marks a sharp \textit{increase} in the false alarm rate's derivative, yielding an elevated FAR at high SNR.
Thus the sharp drops in the left-hand side of Fig. \ref{many_models} should actually be understood as conservative.}

Although we have no formal criteria with which to assess our various background models, it is informative to compare the implications of these models for $P(S)$.
Shown in Fig. \ref{pAstro_figure} are the inferred probabilities $P(S)$ assigned to loud triggers under each background model, following Eq. \eqref{pAstro}, as a function of their SNR.
As seen in Fig. \ref{many_models}, the Gaussian KDE model decreases exponentially with $\rho$, falling off far more quickly than the astrophysical signal model.
Hence $P(S)$ increases rapidly with increasing $\rho$ in Fig. \ref{pAstro_figure}, and nearly any signal candidate with $\rho>\rho_\mathrm{thresh}$ would be deemed almost certainly real.
This model is \textit{indefensibly optimistic}, assuming that loud noise events occur with negligible probability.
Any claimed detections based on this noise model would therefore likely be dismissed.
Similarly, the $p_n(\rho)\propto\rho^{-8}$ power-law model falls off more steeply than the signal model, and so louder triggers are considered more likely to be astrophysical.
While this behavior seems reasonable, it is again hard to defend this choice of background model.

The flat ($\alpha=0$) model, in contrast, falls off less steeply than the astrophysical signal distribution.
As seen in Fig. \ref{pAstro_figure}, this leads to the strange conclusion that triggers of increasing $\rho$ are \textit{less likely} to represent true gravitational-wave signals.
This conclusion is problematic.
It suggests that the ranking statistic $\rho$ is a poor measure of a candidate's significance.
If we believe data analysis pipelines and their detection statistics $\rho$ to be reasonably-behaved, we are forced to reject the flat background model.
Similarly, we should reject \textit{any} background model $p_n(\rho)$ that falls off more shallowly than the signal distribution $p_s(\rho)$.
Background models that are shallower than the expected signal distribution should be characterized as \textit{overly pessimistic}, assuming so many loud noise events that those candidates with the lowest detection statistics are paradoxically the \textit{most likely} to be real.

The conservative choice, then, is a background model that exactly parallels the expected distribution of astrophysical triggers:
	\begin{equation}
	p_n(\rho) \propto p_s(\rho).
	\end{equation}
For the case of \texttt{PyCBC}, this condition leads us to adopt the power-law model $p_n(\rho) \propto \rho^{-4}$ for the noise distribution above our measured background.
The resulting values of $P(S)$ will thus be the \textit{most conservative} lower limits one can place without asserting that louder signals are less significant. 

This choice of background model has the additional property that it exactly recovers Eq. \eqref{naive}, our back-of-the-envelope estimate of $P(S)$ for a gravitational-wave candidate falling above the measured single-detector background.
When assuming $p_n(\rho)\propto p_s(\rho)$, we may rewrite Eq. \eqref{pAstroIntegrand} as
	\begin{equation}
	\begin{aligned}
	P(S|\Lambda_s,\Lambda_n)
		&= \frac{\Lambda_s p_s(\rho)}{\Lambda_s p_s(\rho) + \Lambda_n p_n(\rho)} \\
		&= \frac{\Lambda_s}{\Lambda_s + \frac{p_n(\rho)}{p_s(\rho)} \Lambda_n} \\
		&= \frac{\Lambda_s}{\Lambda_s + \frac{p_n(\rho)}{p_s(\rho)}\Lambda_n } 
			\frac{\int_{\rho_\mathrm{thresh}}^{\rho_\mathrm{max}} p_s(\rho') d\rho'}
				{\int_{\rho_\mathrm{thresh}}^{\rho_\mathrm{max}} p_s(\rho') d\rho'} \\
		&= \frac{\Lambda_s \int_{\rho_\mathrm{thresh}}^{\rho_\mathrm{max}} p_s(\rho') d\rho'}
			{\Lambda_s \int_{\rho_\mathrm{thresh}}^{\rho_\mathrm{max}} p_s(\rho') d\rho' 
				+ \Lambda_n \int_{\rho_\mathrm{thresh}}^{\rho_\mathrm{max}} p_n(\rho') d\rho'} \\
		&= \frac{\lambda_s}{\lambda_s + \lambda_n},
	\end{aligned}
	\label{consPAstro}
	\end{equation}
exactly equal to our naive estimate in Eq. \eqref{naive}.
In moving from the third step of Eq. \eqref{consPAstro} to the fourth, we have used the fact that the ratio $p_n(\rho)/p_s(\rho) = p_n(\rho')/p_s(\rho') = \mathrm{constant}$ above $\rho_\mathrm{thresh}$.

\section{Examples}
\label{sec:examples}

As an example of the above machinery, consider two hypothetical gravitational-wave triggers observed by a single detector.
The first lies within the measured H1 background at $\rho=9$.
Given the H1 single-detector background in Fig. \ref{H1Background}, this trigger would be assigned a false alarm rate $\mathrm{FAR} = 42\,(\pm17)\yr^{-1}$ and a probability $P(S) = 0.04$ of astrophysical origin.

Secondly, consider a single-detector trigger falling beyond the measured background, with $\rho=15$ (approximately the single-detector SNR of the binary black hole GW150914).
Using the conservative background model $p_n(\rho)\propto p_s(\rho)$, we would limit the astrophysical probability of such an event to $P(S)>0.44$.

\begin{figure}
\centering
\includegraphics[width=0.48\textwidth]{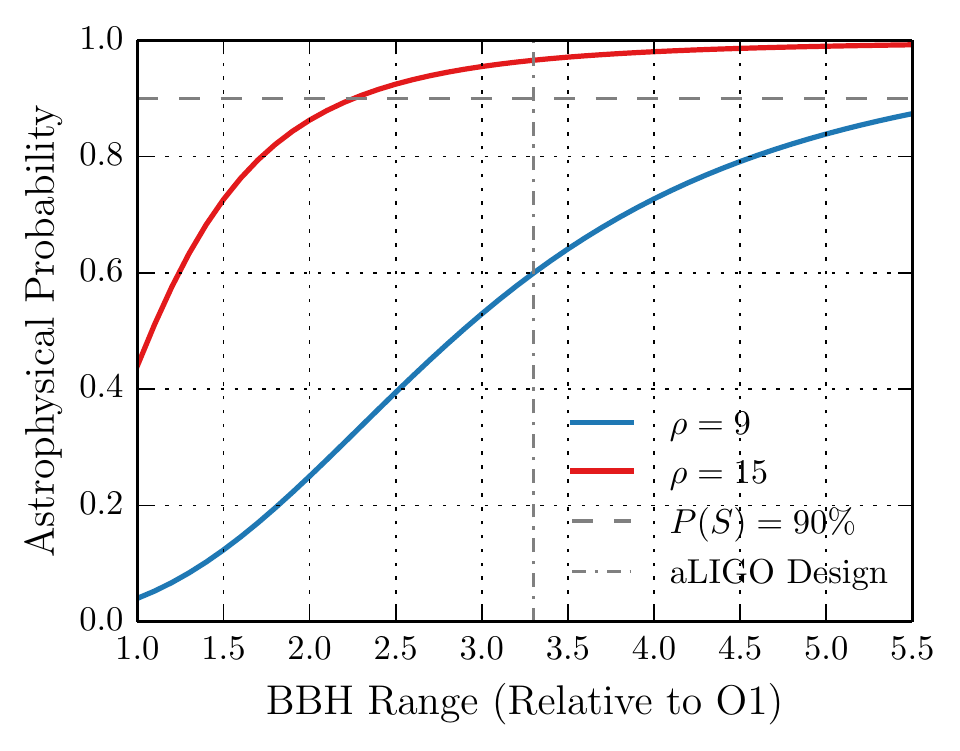} 
\caption{
The probability $P(S)$ of astrophysical origin for events observed with a single Advanced LIGO detector, as a function of the detector's sensitive range.
Ranges are given relative to Advanced LIGO's O1 range.
The blue and red curves indicate $P(S)$ for single-detector candidates with $\rho=9$ and $\rho=15$, respectively. 
For reference, we show Advanced LIGO's expected range improvement for binaries of total mass $40\,M_\odot$ (vertical dot-dashed line).
We additionally mark the $P(S) = 90\%$ confidence threshold (horizontal dashed line), corresponding to the approximate astrophysical probability of the marginal binary black hole candidate LVT151012 \cite{Abbott2016}.
At Advanced LIGO's current sensitivity, single-detector candidates of $\rho=9$ and $\rho=15$ would be assigned astrophysical probabilities of $P(S)=0.04$ and $P(S)>0.44$, respectively.
If Advanced LIGO's range were improved by a factor of three, these candidates would instead have $P(S) = 0.53$ and $P(S)>0.95$.
 }
\label{confidence}
\end{figure}

For fixed $\rho$, the astrophysical probability of a gravitational-wave candidate grows with the expected rate of astrophysical signals.
Hence the astrophysical probabilities of single-detector triggers will increase with the sensitive volume probed by future detectors.
Fig. \ref{confidence}, for instance, demonstrates the expected increase in $P(S)$ as a function of improvement in Advanced LIGO's binary black hole range, relative to its O1 range.
The blue and red curves illustrate $P(S)$ for our hypothetical candidates at $\rho=9$ and $\rho=15$, respectively.
The vertical dot-dashed line marks the expected range improvement for binary black holes of total mass $40\,M_\odot$ (the mean total mass of binary black holes observed thus far) between Advanced LIGO's O1 and design sensitivities.
For reference, the horizontal dashed line marks $P(S)=0.9$, the approximate astrophysical probability of the binary black hole candidate LVT151012 \cite{Abbott2016}.
A marginally-significant signal measured by Advanced LIGO during O1, LVT151012 was observed with a two-detector SNR of $\rho=9.7$, corresponding to a $1.7\sigma$ detection \cite{Abbott2016}.

If Advanced LIGO's range were improved by a factor of three, the $\rho=9$ single-detector candidate (which would have $P(S)=0.04$ in O1) would be assigned a probability $P(S)=0.53$ of astrophysical origin.
While far from a confident detection, even marginally-confident events like this may prove valuable when inferring properties of the binary black hole population (e.g. mass distributions and coalescence rate) \cite{Farr2015}.
The $\rho=15$ candidate, meanwhile, would be assigned $P(S)=0.95$, greater than the confidence assigned to LVT151012.

\section{Discussion}

In this paper, we have explored a practical scheme for assigning detection confidences to compact binary coalescence candidates discovered in times when only one gravitational-wave detector is operating.
Searching for gravitational-wave signals in such times will accelerate the rate of discovery by increasing the effective duty cycle of the current-generation detector network.
The additional live time may yield more detections of binary black holes, potentially with interesting spins or mass ratios.
Additional live time also increases the likelihood of observing the first binary neutron star and/or neutron star-black hole \blue{candidates}; as discussed in Sect. 1, there is a non-negligible chance that such systems will be first observed single-detector time.

A single-detector search is, of course, necessarily less sensitive than a coincident search between two or more gravitational-wave detectors.
Given the example Advanced LIGO O1 background in Fig. \ref{H1Background}, for instance, a single-detector \blue{binary black hole} candidate would require $\rho > 9$ in order to be even marginally identifiable.
In comparison, standard Advanced LIGO searches over coincident data can detect signals with single-detector SNRs of $\rho\sim8$ at $4\sigma$ confidence \cite{Abbott2016}.
The framework discussed here will therefore be most relevant to the loudest gravitational-wave events in single-detector time.

A significant challenge in assigning astrophysical probabilities to very loud events is that the single-detector background is not well measured in this regime.
While we can place an experimental upper limit on the integrated rate $\lambda_n$ of loud background events, the distribution $p_n(\rho)$ of such events is virtually unconstrained.
In Sect. \ref{sec:beyond}, we argued that the noise model $p_n(\rho)\propto p_s(\rho)$ can be used to place conservative lower limits on the astrophysical probability of events falling beyond the measured background.

This model offers several advantages over virtually any other choice.
First, a background model that is any shallower than our proposed choice would imply that louder candidates are \textit{less likely} to be astrophysical.
The model $p_n(\rho)\propto p_s(\rho)$ is therefore the \textit{most conservative} choice one can make that is consistent with both a sensible search pipeline and a well-defined detection statistic $\rho$.
Secondly, this choice is qualitatively consistent with observed power-law distributions of high-SNR ``glitches'' in the Advanced LIGO interferometers \cite{Abbott2016h}.
Finally, the model recovers our naive estimate [Eq. \eqref{naive}] of a trigger's probability of astrophysical origin.
Thus our proposed background distribution is consistent, in a sense, with maximum ignorance of the noise properties above our measured background.
If one were to reject altogether the notion of extrapolating beyond the measured background, one would obtain a limit on $P(S)$ identical to that obtained with our chosen background model.

Using Advanced LIGO's H1 detector as an example, we found in Sect. \ref{sec:examples} that loud binary black hole candidates in a single detector can be assigned probabilities $P(S)\gtrsim0.4$ of astrophysical origin with current instruments.
Within the next several years, though, it may be possible to make single-detector binary black hole observations with confidences exceeding $P(S)\gtrsim0.9$.

Although we have specifically focused on binary black hole mergers, the same methodology can be straightforwardly extended to other gravitational-wave sources like binary neutron stars.
In the binary black hole case, our prior $p(\Lambda_s)$ on the astrophysical signal rate was informed by direct Advanced LIGO measurements.
In the case of binary neutron star mergers (which have yet to be observed), our prior would instead follow constraints derived from \blue{electromagnetic} binary neutron star observations and population synthesis models \cite{Abadie2010,Kim2015,Vangioni2016}.
\blue{The resulting astrophysical probabilities $P(S)$ will likely be much weaker, given the large uncertainty in the rate of binary neutron star mergers.
Additionally, it may not always be clear which signal class (binary neutron star or low-mass binary black hole) a candidate falls in; there may therefore be some ambiguity in choosing which rate prior to adopt for a given candidate.
However, our method nonetheless offers a means of quantifying the significance of the first loud binary neutron star candidate, should it fall in single-detector time.
}

We note that future commissioning efforts and advances in seismic isolation and interferometer control may further improve detector duty cycles.
The GEO 600 detector, for instance, can sustain duty cycles of up to $\sim90\%$ \cite{2010CQGra..27h4003G}.
Additionally, in the coming years Advanced LIGO will be joined by a host of new detectors, including Advanced Virgo \cite{Acernese2015}, KAGRA \cite{Aso2013}, and LIGO-India \cite{Iyer2011}.
As both the number of operational detectors and their duty cycles grow, we may have reduced need for single-detector analyses.
The operation of a truly global network of detectors, however, raises new and unique challenges, including the coordination of observing runs, maintenance schedules, and commissioning breaks.
The ability to make meaningful observations with single detectors may therefore remain crucial, affording greater commissioning flexibility, increased network duty cycle, and greater opportunities for gravitational-wave discoveries.

\acknowledgements{
We wish to thank the \texttt{PyCBC} development team for use of the O1 trigger list.
We also thank Jolien Creighton, Tom Dent, Reed Essick, Will Farr, Chad Hanna, Alex Nitz, Surabhi Sachdev, and David Shoemaker for helpful comments and conversation, and the anonymous referees for their valuable feedback.
S. D. would like to thank Alan Weinstein and Albert Lazzarini for his visit to LIGO at Caltech and the LIGO Laboratory for the local travel and hospitality.
T. C., J. K., T. M., and A. W. are members of the LIGO Laboratory, supported by funding from the U. S. National Science Foundation.
LIGO was constructed by the California Institute of Technology and Massachusetts Institute of Technology with funding from the National Science Foundation and operates under cooperative agreement Grant No. PHY-0757058.
This paper carries the LIGO Document Number LIGO-P1700032.
}

\begin{appendix}
\section{The \texttt{PyCBC} Search Region (ii)}
\label{bulkAppendix}

As an example of our proposed single-detector method, in Figs. \ref{H1Background} and \ref{many_models} we show results using H1 triggers from search region (ii) of the \texttt{PyCBC} pipeline \cite{Canton:2014ena,Usman2016,pycbc} during Advanced LIGO's O1 observing run \cite{Abbott2016}.
In searches for compact binary coalescences, the measured background distribution varies substantially with respect to different template parameters (like binary masses and spins).
Searches therefore divide the template parameter space into regions defined by similar background properties \cite{Usman2016}.
Region (ii) of the O1 \texttt{PyCBC} search comprises templates with chirp masses $\mathcal{M}>1.74\Msun$ and peak amplitudes at frequencies $f_\mathrm{peak} \geq 100\,\mathrm{Hz}$ \cite{Abbott2016}.
At leading order, the chirp mass $\mathcal{M}$ governs the phase evolution of a compact binary inspiral;
it is defined in terms of a binary's total mass $M$ and symmetric mass ratio $\eta$ by $\mathcal{M} = \eta^{3/5} M$.

\blue{The definition of region (ii) serves to minimize the effects of noise transients, producing a well-behaved search background in this region.}
Note that region (ii) covers the parameter space of all binary black holes observed in O1 \cite{Abbott2016}.
\blue{Other search regions containing short duration and/or high mass templates [the \texttt{PyCBC} search region (iii), for instance] suffer considerably more contamination due to instrumental artifacts \cite{Abbott2016}; it is unclear if the background extrapolation presented in this paper can be applied in such regions.}

Work is currently underway to more accurately model the \texttt{PyCBC} background variation across template parameter space \cite{Nitz_Prep}.
This development will yield more accurate estimates of trigger significances in current and future observing runs.

\section{Sensitive Volume}
\label{sec:volume}

As discussed in Sect. \ref{calc_confidence}, the expected rate of binary black hole signals in single-detector time is given by $\Lambda_s = R \langle V\rangle_\mathrm{single}$, where $R$ is the astrophysical rate density of binary black hole mergers and $\langle V \rangle_\mathrm{single}$ is the population-averaged sensitive volume inside of which signals are expected to have $\rho>\rho_\mathrm{min}$ \cite{Abbott2016f,Abbott2016g}.
While $\langle V\rangle_\mathrm{single}$ should in principle be computed via the numerical injection and recovery of simulated gravitational-wave signals into Advanced LIGO data, we estimate $\langle V\rangle_\mathrm{single}$ by scaling the sensitive time-volume $\langle VT\rangle_\mathrm{coinc}$ presented in Ref. \cite{Abbott2016g}.

Using the \texttt{PyCBC} pipeline and assuming a power-law distribution of black hole masses, the sensitive time-volume of the H1-L1 network was found to be $\langle VT \rangle_\mathrm{coinc} = 0.0154\,\mathrm{Gpc}^{3}\yr$ after 17 days of observation during O1 \cite{Abbott2016g}.
This time-volume estimate assumes a minimum network detection statistic $\rho_\mathrm{min} = 8.0$.
In this paper, we are instead interested in the sensitive volume $\langle V\rangle_\mathrm{single}$ corresponding to $\rho_\mathrm{min}=7.1$ in a single detector.
We will obtain this volume by rescaling the sensitive time-volume from Ref. \cite{Abbott2016g}.

Note that network matched-filter SNRs are obtained by adding single-detector SNRs in quadrature.
The network SNR of a gravitational-wave signal therefore scales as
	\begin{equation}
	\mathrm{SNR} \propto \frac{\sqrt{N}}{D},
	\end{equation}
where $D$ is the distance to the gravitational-wave source and $N$ is the number of detectors comprising the network, assuming approximately equal SNRs in each detector.
Under this scaling law, the population-averaged volume inside of which binary black holes have $\rho>\rho_\mathrm{min}$ in a single detector is approximately
	\begin{equation}
	\langle V \rangle_\mathrm{single} = \frac{1}{2^{3/2}} \left(\frac{8.0}{7.1}\right)^3\frac{ \langle VT \rangle_\mathrm{coinc} }{17\,\mathrm{days}}.
	\label{scaling}
	\end{equation}
The factor of $\left(8.0/7.1\right)^3$ scales the sensitive volume defined with respect to network $\rho_\mathrm{min}=8.0$ to the volume corresponding to network $\rho_\mathrm{min}=7.1$.
The leading factor of $2^{-3/2}$, meanwhile, moves from a two-detector SNR to a single-detector SNR.

\end{appendix}

\end{document}